\documentclass[12pt,a4paper]{article}

\title{Single-charge rotating black holes in four-dimensional gauged supergravity}
\author{David D. K. Chow}
\date{}

\usepackage{amsmath}
\usepackage{amsfonts}
\usepackage{amssymb}
\usepackage{latexsym}

\makeatletter
\@addtoreset{equation}{section}
\makeatother

\newcommand{\be}{\begin{equation}}
\newcommand{\ee}{\end{equation}}
\newcommand{\ben}{\begin{equation}}
\newcommand{\een}{\end{equation}}
\newcommand{\bea}{\setlength\arraycolsep{2pt} \begin{eqnarray}}
\newcommand{\eea}{\end{eqnarray}}

\newcommand{\nnr}{\nonumber \\}
\newcommand{\lbl}{\label}

\newcommand{\eq}[1]{(\ref{#1})}
\newcommand{\se}{\section}
\newcommand{\sse}{\subsection}

\newcommand{\qd}{\quad}

\newcommand{\lt}{\left}
\newcommand{\rt}{\right}
\newcommand{\fr}{\frac}

\newcommand{\tf}{\tfrac}

\newcommand{\wtd}{\widetilde}

\newcommand{\df}{\textrm{d}}
\newcommand{\expe}[1]{\textrm{e}^{#1}}
\newcommand{\na}{\nabla}
\newcommand{\ol}{\overline}
\newcommand{\pd}{\partial}
\newcommand{\ra}{\rightarrow}

\newcommand{\sr}{\sqrt}

\newcommand{\gd}{\delta}
\newcommand{\gD}{\Delta}

\newcommand{\gve}{\varepsilon}

\newcommand{\gq}{\theta}

\newcommand{\gk}{\kappa}

\newcommand{\gr}{\rho}

\newcommand{\gS}{\Sigma}

\newcommand{\gf}{\phi}
\newcommand{\gF}{\Phi}
\newcommand{\gvf}{\varphi}

\newcommand{\gW}{\Omega}

\newcommand{\uU}{\textrm{U}}

\newcommand{\im}{\textrm{i}}

\newcommand{\SO}{\textrm{SO}}

\newcommand{\cL}{\mathcal{L}}

\newcommand{\cN}{\mathcal{N}}
\newcommand{\cO}{\mathcal{O}}

\newcommand{\ds}{\textrm{d} s^2}

\begin{document}

\thispagestyle{empty}

\begin{flushright}
MIFPA-10-50
\end{flushright}
\vspace*{100pt}
\begin{center}
{\bf \Large{Single-charge rotating black holes in four-dimensional gauged supergravity}}\\
\vspace{50pt}
\large{David D. K. Chow}
\end{center}

\begin{center}
\textit{George P. \& Cynthia W. Mitchell Institute for Fundamental Physics \& Astronomy,\\
Texas A\&M University, College Station, TX 77843-4242, USA}\\
{\tt chow@physics.tamu.edu}\\
\vspace{30pt}
{\bf Abstract\\}
\end{center}
We consider four-dimensional $\uU (1)^4$ gauged supergravity, and obtain asymptotically AdS$_4$, non-extremal, charged, rotating black holes with one non-zero $\uU (1)$ charge.  The thermodynamic quantities are computed.  We obtain a generalization that includes a NUT parameter.  The general solution has a discrete symmetry involving inversion of the rotation parameter, and has a string frame metric that admits a rank-2 Killing--St\"{a}ckel tensor.

\newpage


\se{Introduction}


Over the past 20 years, there has been a substantial effort to construct black hole solutions of supergravity theories, with increasing complexity.  Before the advent of the AdS/CFT correspondence, such efforts were largely directed at asymptotically flat black holes of ungauged supergravities, but have since turned to asymptotically AdS black holes of gauged supergravities.  A large variety of solutions are now known, in various dimensions, carrying various numbers of independent angular momenta and $\uU (1)$ charges.

In ungauged supergravities, one can start from an uncharged solution, and then introduce charges in an algorithmic fashion by solution generating techniques.  The starting point for these is typically the Myers--Perry solution \cite{myeper}, which generalizes the Kerr solution from $D = 4$ spacetime dimensions to higher dimensions.  This results in the 4-charge Cveti\v{c}--Youm solution in $D = 4$ \cite{cveyou, chcvlupo}, the 3-charge Cveti\v{c}--Youm solution in $D = 5$ \cite{cveyou2}, and the 2-charge Cveti\v{c}--Youm solution in $D \geq 6$ \cite{cveyou3}.  Another starting point is the doubly rotating black ring in $D = 5$ \cite{pomsen}, which has a 3-charge generalization \cite{galsch}.  These examples possess the maximum number of independent angular momenta and $\uU (1)$ charges.

In gauged supergravities, there is a potential in the Lagrangian that inhibits such solution generating techniques.  One must therefore resort to guesswork, aided by the other known solutions.  For supersymmetric solutions, rotation is required, which makes the guesswork challenging.  In the simplest case of Einstein gravity with a cosmological constant, the Kerr--AdS solution in $D = 4$ has long been known \cite{carter0, carter}, and since generalized to higher dimensions \cite{hahuta, gilupapo, gilupapo2}.  We now have a large catalogue of asymptotically AdS charged and rotating black hole solutions of gauged supergravities in $D = 4, 5, 6, 7$; see \cite{emprea, chow} for a review of these.  However, we still do not know black hole solutions of gauged supergravity in $D = 4, 5, 7$ with the maximum number of independent angular momenta and $\uU (1)$ charges.  Only in the simpler case of $\uU (1)$ gauged supergravity in $D = 6$ is such a programme complete \cite{chow2}.  The known solutions can appear rather complicated, but a hint that more general solutions should be amenable to guesswork is that all known solutions have string frame metrics that possess rank-2 Killing--St\"{a}ckel tensors \cite{chow}.  This suggests some underlying structure to the geometries that is yet to be fully understood.

In the hope of taming the contents of this bestiary of solutions, in this paper we turn to arguably the most basic, but not the simplest, example of a charged and rotating asymptotically AdS black hole solution of gauged supergravity.  We work in 4 dimensions, where there is an $\cN = 8$ gauged supergravity theory with gauge group $\SO (8)$, whose Cartan subgroup is $\uU (1)^4$.  The previously discovered solutions of the theory are: static AdS black holes with 4 independent $\uU (1)$ charges \cite{dufliu}; and rotating AdS black holes with the 4 $\uU(1)$ charges pairwise equal \cite{chcvlupo}, which includes the Kerr--Newman--AdS solution \cite{carter0, carter} of the Einstein--Maxwell system (the dyonic Kerr--Newman--AdS solution was given explicitly in \cite{plebanski}).

We shall write down a rotating solution for which only 1 of the 4 $\uU (1)$ charges is non-zero.  We compute the thermodynamical quantities and find that there are no supersymmetric solutions.  Then, we generalize the solution to include a NUT parameter, and can perform electric/magnetic duality.  Like the uncharged solution, there is a discrete symmetry that includes the rotation parameter being inverted through the AdS radius.  We write down a rank-2 Killing St\"{a}ckel tensor for the string frame metric, which implies separability of the string frame Hamilton--Jacobi equation for geodesic motion and of the Klein--Gordon equation.


\se{AdS$_4$ black hole solution}


4-dimensional $\uU (1)^4$ gauged supergravity is a consistent truncation of the maximal $\cN = 8$, $\SO (8)$ gauged supergravity.  It is $\cN = 2$ supergravity coupled to 3 abelian vector multiplets.  The full bosonic Lagrangian for $\uU (1)^4$ gauged supergravity was given in \cite{cveticetal}.  Without axions, which suffices for our purposes here, the truncated bosonic Lagrangian was given in \cite{dufliu}.

We shall further truncate and consider black hole solutions with a single $\uU (1)$ charge.  The bosonic fields of this truncation are a graviton, a vector and a scalar, and the Lagrangian is
\ben
\cL_4 = R \star 1 - \tf{3}{2} \star \df \gvf \wedge \df \gvf - \tf{1}{2} \expe{3 \gvf} \star F_{(2)} \wedge F_{(2)} + 3 g^2 (\expe{\gvf} + \expe{- \gvf}) \star 1 ,
\een
where $F_{(2)} = \df A_{(1)}$.  We have used a non-canonical normalization for the scalar kinetic term to avoid awkward $\sr{3}$ factors from appearing.  The resulting field equations are
\bea
&& G_{a b} = \tf{3}{2} (\na_a \gvf \, \na_b \gvf - \tf{1}{2} \, \na^c \gvf \, \na_c \gvf \, g_{a b}) + \tf{1}{2} \expe{3 \gvf} (F{_a}{^c} F_{b c} - \tf{1}{4} F^{c d} F_{c d} g_{a b}) + \tf{3}{2} g^2 (\expe{\gvf} + \expe{-\gvf}) g_{a b} , \nnr
&& \na_a (\expe{3 \gvf} F^{a b}) = 0 , \nnr
&& \square \gvf - \tf{1}{4} \expe{3 \gvf} F^{a b} F_{a b} + g^2 (\expe{\gvf} - \expe{- \gvf}) = 0 . \lbl{fieldeq}
\eea

A charged and rotating black hole solution is
\bea
\ds & = & \fr{1}{\sr{H} \gr^2 \Xi^2} \bigg( - \gD_\gq (\gD_\gq \gD_r - V_r^2 a^2 \sin^2 \gq) \, \df t^2 - 2 m r c \sr{1 + a^2 g^2 s^2} \gD_\gq a \sin^2 \gq \, 2 \, \df t \, \df \gf \nnr
&& + (\gD_\gq \wtd{V}_r^2 a^2 - \gD_r \sin^2 \gq) a^2 \sin^2 \gq \, \df \gf^2 \bigg) + \sr{H} \lt( \fr{\gr^2}{\gD_r} \, \df r^2 + \fr{\gr^2}{\gD_\gq} \, \df \gq^2 \rt) , \nnr
A_{(1)} & = & \fr{2 m r s}{H \gr^2} \lt( c \gD_\gq \, \fr{\df t}{\Xi} - a \sr{1 + a^2 g^2 s^2} \sin^2 \gq \, \fr{\df \gf}{\Xi} \rt) , \nnr
\gvf & = & \tf{1}{2} \log H , \lbl{solution}
\eea
where
\bea
\gD_r & = & r^2 + a^2 - 2 m r + g^2 r^2 (r^2 + 2 m s^2 r + a^2) , \qd \gD_\gq = 1 - a^2 g^2 \cos^2 \gq , \qd \gr^2 = r^2 + a^2 \cos^2 \gq , \nnr
V_r^2 & = & (1 + g^2 r^2) (1 + g^2 r^2 + 2 m s^2 r g^2) , \qd \wtd{V}_r^2 = (1 + r^2 / a^2) (1 + r^2 / a^2 + 2 m s^2 r / a^2) , \nnr
H & = & 1 + \fr{2 m r s^2}{\gr^2} , \qd s = \sinh \gd , \qd c = \cosh \gd , \qd \Xi = 1 - a^2 g^2 .
\eea

The Boyer--Lindquist-type coordinates used are asymptotically static.  If $m = 0$, then the coordinate change
\ben
\Xi \hat{r}^2 \sin^2 \hat{\gq} = (r^2 + a^2) \sin^2 \gq , \qd \hat{r}^2 \cos^2 \hat{\gq} = r^2 \cos^2 \gq ,
\een
gives
\ben
\ds = - (1 + g^2 \hat{r}^2) \, \df t^2 + \fr{\df \hat{r}^2}{1 + g^2 \hat{r}^2} + \hat{r}^2 \, \df \hat{\gq}^2 + \sin^2 \hat{\gq} \, \df \gf^2 .
\een
This is simply anti-de Sitter spacetime, and we see that $t$ and $\gf$ are canonically normalized.

The solution has 4 parameters: a mass parameter $m$; a rotation parameter $a$; a charge parameter $\gd$; and a gauge-coupling constant $g$.  In the absence of rotation, with $a = 0$, the solution reduces to a particular case of the 4-charge static solution \cite{dufliu}, but with only 1 of the 4 charges non-zero.  In the absence of charge, with $\gd = 0$, the solution reduces to the 4-dimensional Kerr--AdS metric \cite{carter0, carter}.  In the absence of gauging, with $\gd = 0$, the solution reduces to the 4-charge Cveti\v{c}--Youm solution \cite{cveyou, chcvlupo}, but with only 1 of the 4 charges non-zero.

To find this solution, we have been helped by the structure of these limits.  We have also been helped by the structure of the black hole solution in 5-dimensional gauged supergravity carrying a single non-zero rotation parameter and a single non-zero $\uU (1)$ charge \cite{chcvlupo2}.


\se{Thermodynamics}


The outer black hole horizon is located at the largest root of $\gD_r (r)$, say at $r = r_+$.  Its angular velocity $\gW$ is constant over the horizon and is obtained from the Killing vector
\ben
l = \fr{\pd}{\pd t} + \gW \fr{\pd}{\pd \gf}
\een
that becomes null on the horizon.  The angular momentum $J$ is given by the Komar integral
\ben
J = \fr{1}{16 \pi} \int _{S_\infty^2} \! \star \df K ,
\een
where $K$ is the 1-form obtained from the Killing vector $\pd / \pd \gf$.  The electrostatic potential, which is also constant over the horizon, is $\gF = l \cdot A_{(1)} | _{r = r_+}$.  The conserved electric charge is
\ben
Q = \fr{1}{16 \pi} \int_{S_\infty^2} \! \expe{3 \gvf} \star F_{(2)} .
\een
The horizon area $A$ is obtained by integrating the square root of the determinant of the induced metric on a time slice of the horizon.  The surface gravity $\gk$, again constant over the horizon, is given by $l^b \na_b l^a = \gk l^a$ evaluated on the horizon.  As usual, we take the temperature to be $T = \gk / 2 \pi$ and the entropy to be $S = A / 4$.

One finds that $T \, \df S + \gW \, \df J + \gF \, \df Q$ is an exact differential, and so we may integrate the first law of black hole mechanics,
\ben
\df E = T \, \df S + \gW \, \df J + \gF \, \df Q ,
\een
to obtain an expression for the thermodynamic mass $E$.

There are several other definitions of mass for asymptotically AdS spacetimes in the literature.  The AMD (Ashtekar--Magnon--Das) mass is one such definition, for 4 dimensions \cite{ashmag} and higher \cite{ashdas}.  One introduces a conformally rescaled metric $\ol{g}_{a b} = \gW^2 g_{a b}$, with $\gW = 0$ and $\df \gW \neq 0$ on the conformal boundary.  Its Weyl tensor is $\ol{C}{^a}{_{b c d}}$, and we define $\ol{n}_a = \pd_a \gW$.  For an asymptotic Killing vector field $K$, which here is $K = \pd / \pd t$, there is an associated conserved quantity.  In 4 dimensions, the AMD mass is
\ben
E = \fr{1}{8 \pi g^3} \int_\gS \! \df \ol{\gS}_a \, \gW \ol{n}^c \ol{n}^d \ol{C}{^a}{_{c b d}} K^b , \lbl{AMDmass}
\een
where $\df \ol{\gS}_a$ is the area element of the $S^2$ section of the conformal boundary.  The AMD mass has previously been used to compute the mass of various asymptotically AdS rotating black holes \cite{gipepo, kunluc, chlupo}.  For definiteness, we take $\gW = 1 / g r$ for our solution.  As $r \ra \infty$, one finds that the Weyl tensor component $C{^t}{_{r t r}}$ behaves as
\ben
C{^t}{_{r t r}} = \fr{m [2 + (1 + a^2 g^2) s^2]}{2 \Xi g^2 r^5} (2 \Xi + a^2 g^2 \sin^2 \gq) + \cO \lt( \fr{1}{r^6} \rt) .
\een
The conformal boundary has metric
\ben
\df \ol{s}_3^2 = - \fr{\gD_\gq}{\Xi} \, \df t^2 + \fr{1}{g^2 \gD_\gq} \, \df \gq^2 + \fr{\sin^2 \gq}{\Xi g^2} \, \df \gf^2 .
\een
Substituting these into \eq{AMDmass}, we can compute the AMD mass, and find that it agrees with the thermodynamic mass.

In summary, we find the thermodynamic quantities
\bea
E & = & \fr{m [2 + (1 + a^2 g^2) s^2]}{2 \Xi^2} , \nnr
S & = & \fr{\pi \sr{(r_+^2 + a^2) (r_+^2 + a^2 + 2 m s^2 r_+)}}{\Xi} , \qd T = \fr{r_+^2 - a^2 + g^2 r_+^2 (3 r_+^2 + a^2 + 4 m s^2 r_+)}{4 \pi r_+ \sr{(r_+^2 + a^2) (r_+^2 + a^2 + 2 m s^2 r_+)}} , \nnr
J & = & \fr{\sr{1 + a^2 g^2 s^2} c m a}{\Xi^2} , \qd \gW = \fr{\sr{1 + a^2 g^2 s^2} a (1 + g^2 r_+^2)}{c (r_+^2 + a^2)} , \nnr
Q & = & \fr{m s c}{2 \Xi} , \qd \gF = \fr{2 m s c r_+}{r_+^2 + a^2 + 2 m s^2 r_+} .
\eea

Supersymmetric AdS$_4$ black holes are known \cite{kosper, cvgilupo}.  In this single-charge case, the BPS condition, up to choices of signs, is \cite{cvgilupo}
\ben
E - g J - Q = 0 .
\een
This leads to the relation
\ben
(2 \expe{2 \gd} + 1) (\expe{2 \gd} - 1)^2 a^2 g^2 + (3 \expe{2 \gd} + 1)^2 = 0 ,
\een
which has no solutions.  Hence, there are no supersymmetric solutions with a single $\uU (1)$ charge.


\se{NUT parameter generalization}


A more general solution that includes a NUT parameter $\ell$ is
\bea
\ds & = & \fr{1}{\sr{H} (r^2 + y^2) \Xi^2} \bigg( - (V_y^2 R - V_r^2 Y) \, \df t^2 - \fr{2 c \sr{1 + a^2 g^2 s^2} (m r Y + \ell y R)}{a} 2 \, \df t \, \df \gf \nnr
&& + (\wtd{V}_r^2 Y - \wtd{V}_y^2 R) a^2 \, \df \gf^2 \bigg) + \sr{H} \lt( \fr{r^2 + y^2}{R} \, \df r^2 + \fr{r^2 + y^2}{Y} \, \df y^2 \rt) , \nnr
A_{(1)} & = & \fr{2 m r s}{H (r^2 + y^2)} \lt( c (1 - g^2 y^2) \, \fr{\df t}{\Xi} - \fr{\sr{1 + a^2 g^2 s^2} (a^2 - y^2)}{a} \, \fr{\df \gf}{\Xi} \rt) \nnr
&& + \fr{2 \ell y s}{H (r^2 + y^2)} \lt( c (1 + g^2 r^2) \, \fr{\df t}{\Xi} - \fr{\sr{1 + a^2 g^2 s^2} (r^2 + a^2)}{a} \, \fr{\df \gf}{\Xi} \rt) , \nnr
\gvf & = & \tf{1}{2} \log H , \lbl{NUTsol}
\eea
where
\bea
R & = & r^2 + a^2 - 2 m r + g^2 r^2 (r^2 + 2 m s^2 r + a^2) , \qd Y = a^2 - y^2 + 2 \ell y + g^2 y^2 (y^2 + 2 \ell s^2 y - a^2) , \nnr
V_r^2 & = & (1 + g^2 r^2) (1 + g^2 r^2 + 2 m s^2 r g^2) , \qd \wtd{V}_r^2 = (1 + r^2 / a^2) (1 + r^2 / a^2 + 2 m s^2 r / a^2) , \nnr
V_y^2 & = & (1 - g^2 y^2) (1 - g^2 y^2 - 2 \ell s^2 y g^2) , \qd \wtd{V}_y^2 = (1 - y^2 / a^2) (1 - y^2 / a^2 - 2 \ell s^2 y / a^2) , \nnr
H & = & 1 + \fr{2 (m r + \ell y) s^2}{r^2 + y^2} , \qd s = \sinh \gd , \qd c = \cosh \gd , \qd \Xi = 1 - a^2 g^2 .
\eea
Without any NUT parameter, making the coordinate change $y = a \cos \gq$ and renaming $R(r)$ as $\gD_r (r)$ recovers the AdS$_4$ black hole \eq{solution}.  

We have been guided towards this more general solution, from the solution without any NUT parameter, by symmetry.  In particular, if we let $x = \im r$ and replace $m \ra - \im m$, then the solution is symmetric under the simultaneous interchange of $\ell$ and $m$ and of $x$ and $y$.  We also note that the metric determinant has a very simple expression, with
\ben
\sr{-g} = \fr{\sr{H} (r^2 + y^2)}{a \Xi} .
\een

In the absence of charge, with $\gd = 0$, the solution reduces to the 4-dimensional Kerr--NUT--AdS metric \cite{carter0, carter}.  In the absence of gauging, with $\gd = 0$, the solution reduces to the 4-charge Cveti\v{c}--Youm solution generalized to include a NUT parameter \cite{chcvlupo}, but with only 1 of the 4 charges non-zero; this 1-charge solution has also been rederived by a different method \cite{alcede}.


\se{Symmetries}



\sse{Electric/magnetic duality}


The field equations \eq{fieldeq} admit the electric/magnetic duality symmetry
\ben
F_{(2)} \ra \expe{3 \gvf} \star F_{(2)} , \qd \gvf \ra - \gvf .
\een
We can perform this transformation on the general solution \eq{NUTsol}.  Using the orientation $\gve_{t r y \gf} = 1$, the vector and scalar become
\bea
A_{(1)} & = & \fr{2 m y s}{r^2 + y^2} \lt( \sr{1 + a^2 g^2 s^2} (1 + g^2 r^2) \, \fr{\df t}{\Xi} - \fr{c (r^2 + a^2)}{a} \, \fr{\df \gf}{\Xi} \rt) \nnr
&& - \fr{2 \ell r s}{r^2 + y^2} \lt( \sr{1 + a^2 g^2 s^2} (1 - g^2 y^2) \, \fr{\df t}{\Xi} - \fr{c (a^2 - y^2)}{a} \, \fr{\df \gf}{\Xi} \rt)  , \nnr
\gvf & = & - \tf{1}{2} \log H .
\eea
As before, we recover solutions in \cite{cveyou, chcvlupo} for $g = 0$, in \cite{dufliu} for $a = 0$, and in \cite{carter0, carter, plebanski} for $\gd = 0$.


\sse{Inversion symmetry}


The Kerr--NUT--AdS solutions \cite{chlupo2} of Einstein gravity in arbitrary dimensions possess discrete inversion symmetries, under which a rotation parameter $a_i$ is inverted through the AdS radius $1 / g$.  This type of symmetry had previously been noted in 5 dimensions \cite{chlupo3}.  A metric with over-rotation, $| a_i g | > 1$, is mapped to a metric with under-rotation, $| a_i g | < 1$, under such a symmetry.

The inversion symmetry persists for the general solution with a single $\uU (1)$ charge above, not only for the metric, but for all fields.  Under the transformation
\bea
a \ra \fr{1}{a g^2} , \qd r \ra \fr{r}{a g} , \qd y \ra \fr{y}{a g} , \qd m \ra \fr{m}{a^3 g^3} , \qd \ell \ra \fr{\ell}{a^3 g^3} , \qd \gf \ra g t , \qd g t \ra \gf , \qd s \ra a g s . \nnr
\eea
the solution is invariant.  The peculiar-looking factors of $\sr{1 + a^2 g^2 s^2}$ that appear in the solution are interchanged with $c = \sr{1 + s^2}$ under the inversion transformation.


\sse{Killing tensor}


A general feature of charged and rotating supergravity black hole solutions that generalize the Kerr or Myers--Perry solution seems to be that their string frame metrics admit rank-2 Killing--St\"{a}ckel tensors \cite{chow}.  These are symmetric tensors $K_{a b}$ that satisfy $\na_{( a} K_{b c )} = 0$.

Consider here the string frame metric $\df \wtd{s}^2$, which is related to the original Einstein frame metric $\ds$ by $\ds = \sr{H} \, \df \wtd{s}^2$.  The inverse string frame metric is, including the NUT parameter,
\bea
\lt( \fr{\pd}{\pd \wtd{s}} \rt) ^2 & = & - \fr{a^4}{r^2 + y^2} \lt( \fr{\wtd{V}_r^2}{R} - \fr{\wtd{V}_y^2}{Y} \rt) \, \pd_t^2 - \fr{2 c \sr{1 + a^2 g^2 s^2}}{(r^2 + y^2)} \lt( \fr{m r}{R} + \fr{\ell y}{Y} \rt) \, 2 a \, \pd_t \, \pd_\gf \nnr
&& + \fr{1}{(r^2 + y^2)} \lt( \fr{V_y^2}{Y} - \fr{V_r^2}{R} \rt) a^2 \, \pd_\gf^2 + \fr{R}{r^2 + y^2} \, \pd_r^2 + \fr{Y}{r^2 + y^2} \, \pd_y^2 .
\eea
The components $(r^2 + y^2) \wtd{g}^{a b}$ are additively separable as a function of $r$ plus a function of $y$.

For the string frame metric, a rank-2 Killing--St\"{a}ckel tensor is given by
\ben
\wtd{K} = \wtd{K}^{a b} \, \pd_a \, \pd_b = \fr{a^4 \wtd{V}_y^2}{Y} \, \pd_t^2 - \fr{2 c \sr{1 + a^2 g^2 s^2} \ell y a}{Y} \, 2 \, \pd_t \, \pd_\gf + \fr{V_y^2 a^2}{Y} \, \pd_\gf^2 + Y \, \pd_y^2 - y^2 \lt( \fr{\pd}{\pd \wtd{s}} \rt) ^2 ,
\een
and there is a separability structure.  This induces a rank-2 conformal Killing--St\"{a}ckel tensor, with components $\wtd{K}^{a b}$, for the Einstein frame metric.  The Hamilton--Jacobi equation for geodesic motion and the massive Klein--Gordon equation separate for the string frame metric, whereas only the Hamilton--Jacobi equation for null geodesics and the massless Klein--Gordon equation separate for the Einstein frame metric.


\section{Conclusion}


We have presented an asymptotically AdS rotating black hole solution of 4-dimensional $\uU (1)^4$ gauged supergravity that possesses 1 non-zero $\uU (1)$ charge, and studied some of its properties.  It is arguably the most basic of the charged and rotating AdS black hole solutions of gauged supergravity.  Therefore, we hope that it will provide further insights into constructing more general solutions with the maximum number of independent angular momenta and $\uU (1)$ charges in not only four dimensions, but also higher dimensions.  Such gravitational backgrounds would be useful for studying the AdS/CFT correspondence.




\end{document}